# Multi-island single-electron devices from self-assembled colloidal nanocrystal chains

Dirk N. Weiss[1,3], Xavier Brokmann[1], Laurie E. Calvet[2,4], Marc A. Kastner[2], Moungi G. Bawendi[1]

[1]*Department of Chemistry and* [2]*Department of Physics, Massachusetts Institute of Technology, Cambridge, MA 02139*
[3]*Present address: United Technologies Research Center, 411 Silver Lane, East Hartford, CT 06108*
[4]*Present address: Unité Mixte de Physique CNRS/Thales, TRT, and Université Paris-Sud, F-91767 Palaiseau*

We report the fabrication of multi-island single-electron devices made by lithographic contacting of self-assembled alkanethiol-coated gold nanocrystals. The advantages of this method, which bridges the dimensional gap between lithographic and NC sizes, are (1) that all tunnel junctions are defined by self-assembly rather than lithography and (2) that the ratio of gate capacitance to total capacitance is high. The rich electronic behavior of a double-island device, measured at 4.2 K, is predicted in detail by combining finite element and Monte Carlo simulations with the standard theory of Coulomb blockade with very few adjustable parameters.

The confinement of electrons to small metallic islands leads to the quantization of charge because of the energy barrier associated with adding an extra electron to a small capacitance.[1] As a result, when multiple islands are connected in series, electrons pass through them one at a time, making it possible to create current standards, potentially useful for metrology.[2-4] Although they are usually made by lithography, single-electron transistors (SETs), consisting of individual islands between metallic leads, have also been produced using colloidal nanocrystals (NCs). In principle, the latter could have much smaller capacitances than lithographic islands, resulting in larger Coulomb charging energies, larger operating voltages, higher operating temperatures and more accurate current standards. However, the fabrication of devices with NC islands has been limited to the fortuitous trapping of single or multiple NCs between pre-fabricated electrode gaps.[5-7] In addition, the capacitances between the lithographically-defined electrodes and the NCs are relatively large, eliminating the advantages of the small NC size.

Here we demonstrate a fabrication scheme which bridges the dimensional gap between lithographic and NC sizes, by lithographic contacting of previously self-assembled NC chains. Because one or more NCs are incorporated into the edge of the large electrode, all of the important tunnel junctions are defined by self-assembly rather than lithography. This method allows the fabrication of one-dimensional island arrays, similar to those used for metrology, with predictable electronic characteristics. Specifically, we show that the electronic behaviour of a double-island device can be fully explained using the standard theory of Coulomb blockade with very few adjustable parameters.

Our fabrication consisted of three steps: (1) The 50-nm diameter commercially obtained gold NCs were coated with alkanethiol spacer molecules; the NCs were supplied with charged citrate on their surfaces to keep them suspended in solution. (2) The NCs self assembled into a variety of structures and were deposited on a substrate consisting of $SiO_2$ thermally grown on a degenerate Si crystal; the Si substrate is used as a gate electrode. (3) After locating the self-assembled structure of interest, electron-beam lithography was used to contact the structure with gold source and drain electrodes.

After coating the NCs with octanethiol,[8] degenerately doped Si substrates with 300 nm thermal oxide and alignment markers were placed into the NC solution and left for 15 min, during which NCs precipitated on the substrate. The substrate was then removed and rinsed for 1 min with ethanol and blow-dried with nitrogen. Samples for TEM were produced in the same way on carbon-coated Cu grids. The deposited NCs were imaged with SEM and self-assembled chains were contacted using the following steps: Electron-beam lithography [950,000 molecular weight poly(methylmethacrylate), single resist layer]; development of the exposed resist; brief oxygen-plasma etching to remove organic residues; thermal evaporation of 40 nm Au; and liftoff.

The deposition process results in a variety of structures. In addition to chains [Fig. 1(a)] and branched structures, both indicative of diffusion-limited aggregation,[9] we also observe two- and three-dimensional NC aggregates. Aggregation likely occurs in solution after addition of the thiols in step 1, as indicated by a color change from red to purple.[10] Similar to pyridine, described in previous work,[11] thiols displace the charged citrate molecules from the NC surface, thereby lowering the repulsive force between crystals and inducing agglomeration. The local structure of the chains is often characterized by parallel NC facets [Fig. 1(a) arrows]. The gaps between NCs are 1 nm wide, as seen in TEM. Fig. 1(b) shows a double-NC device with lithographic source and drain electrodes attached. We emphasize that the self-assembled chains can be robust enough to survive all of the steps required for e-beam lithography of the leads. Two NCs framing the two island NCs become part of the electrodes. With this method, devices with N islands generally consist of chains of at least N + 2 NCs. We have measured a low contact resistance for 50-nm wide Au thin-film test stripes on the same substrate contacted with the same method, and infer that the resistance between the electrodes and the incorporated NCs is small. Therefore, the tunnel junctions that determine the device characteristics are all defined by self-assembly.

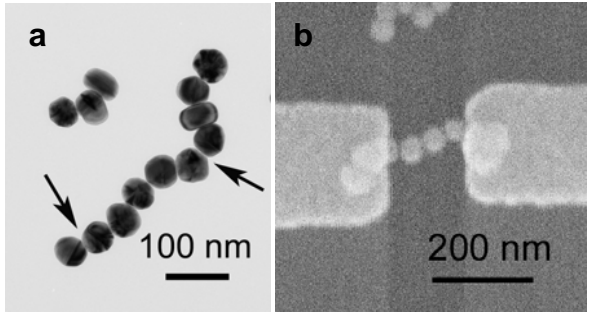

FIG. 1. (a) Transmission-electron microscope (TEM) micrograph of a self-assembled chain of 50-nm Au NCs coated with octanethiol molecules; arrows indicate parallel crystal facets. (b) Double-island device fabricated by contacting a self-assembled chain of NCs. The two NCs framing the two island-particles form part of the electrodes, such that all three tunnel junctions are formed by self assembly and not by lithography.

The samples were held in liquid He and a small (870 μV) AC voltage at frequency 13 Hz was applied between source and drain, together with a DC drain-source voltage. The AC current was measured with a current amplifier and lock-in amplifier. Twenty percent of the devices had resistances of 0.1 – 20 GOhm at room temperature; the remaining 80% were either shorted (probably due to the sintering of NCs during plasma etching) or had infinite resistance. Fifty percent of the conductive devices exhibited Coulomb-blockade at 4.2 K.

Fig. 2(a) shows the differential conductance $dI/dV_{DS}$, as a function of drain-source $V_{DS}$ and gate $V_G$ voltages, of the double-island structure whose micrograph is shown in Fig. 1(b). Near $V_{DS}=0$, the conductance is suppressed because of the energy cost to add an electron to one of the islands. This is called Coulomb blockade. Two Coulomb blockade regions (large and small white diamonds near $V_{DS}=0$) are seen. One also observes four peaks in $dI/dV_{DS}$ (diagonal dark lines) per gate period. The Coulomb charging energy, based on the size of the diamonds is ~20 meV, which is why such well-resolved structure can be seen in $dI/dV_{DS}$ at liquid He temperature.

These data are compared to a simulation of the conductivity plot expected for a double-island device made of four 50-nm diameter spheres with 1 nm dielectric cap layers, separated by 1 nm, connected in series with the outer two incorporated into the electrodes [Fig. 2(b)]. The simulation consists of two parts. First, a finite element calculation is made of the capacitance matrix of the model structure in Fig. 2(b).[12] Second, a semi-classical Monte Carlo simulation is made of the tunneling rate as a function of $V_{DS}$ and $V_G$.[13] In the Monte Carlo simulation, the double island is modeled as a purely classical network of resistors and capacitances [Fig. 2(c)], but charge quantization is imposed. The interdot thiol spacing layers are considered as tunneling barriers that couple each of the two outermost islands to the leads with capacitance $C_L$ and the two centre islands together with capacitance $C_M$. The capacitance between each of the islands and the degenerate silicon substrate is $C_G$.

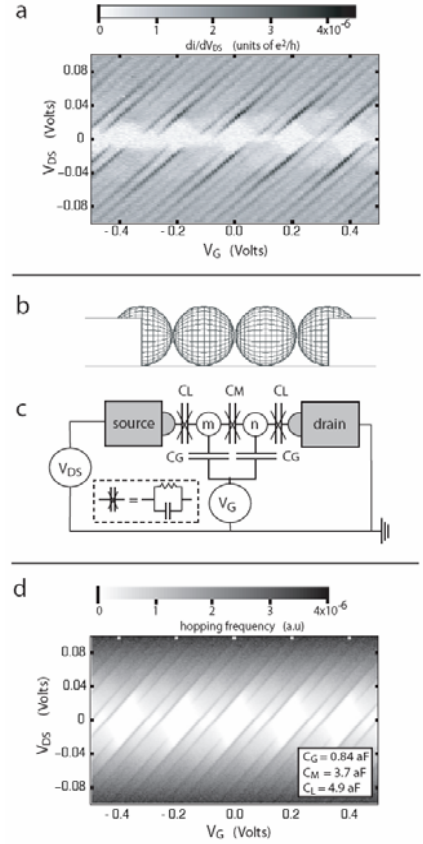

FIG. 2. Experimental and predicted differential conductance plots of the double-island device of Fig. 1(b). (a) Differential conductance measured at 4.2 K; four peaks are found per gate period. Above the threshold for Coulomb Blockade, the current can be described as linear with small oscillations superposed, which give the peaks in $dI/dV_{DS}$. The linear component corresponds to a resistance of ~20GΩ. (b) Ideal NC chain used for finite element calculations. The two framing spheres are ohmically connected to the bar electrodes. (c) Electrical modeling of the device. The silicon substrate functions as a common gate electrode for both islands. (d) Monte Carlo simulation of a stability plot for the double-island device at 4.2 K with capacitance values obtained from finite-element modeling: $C_G = 0.84$ aF (island-gate capacitance), $C_M = 3.7$ aF (inter-island capacitance), $C_L = 4.9$ aF (lead-island capacitance); the resistances are set to 0.1, 10 and 10 GΩ to reproduce the experimental data.

As seen in Fig. 2(d), and as discussed further below, our simulations predict the experimental behaviour quite well. Specifically, the simulation reproduces the two diamonds and the four peaks per gate period. The resistances of the tunnel junctions are adjusted to describe the experimental data. One resistance has to be much smaller than other two in order to give the prominent peaks in $dI/dV_{DS}$; the sum of the resistances is determined directly from the observed average conductance above the threshold for Coulomb blockade. It is remarkable that the simulation provides an excellent description of the data with only the resistances as adjustable parameters.

Deeper understanding of the pattern of the differential conductance is obtained by considering the charge states of the double-island system. These states are labeled by the number of extra electrons on the first and second islands, m and n, respectively. At zero $V_{DS}$ the current is zero unless the chemical potentials of the two islands $\mu_1$ and $\mu_2$ are the same, which is very unlikely at any value of $V_G$. However, at finite $V_{DS}$ $\mu_1$ and $\mu_2$ can both fall between the chemical potentials of drain and source. In particular, electrons can pass through the device by means of a cycle such as (m,n)→(m+1,n)→(m,n+1)→(m,n), and holes can pass by means of a cycle such as (m+1,n+1)→(m+1,n)→(m,n+1)→(m+1,n+1). A step-like increase in the current occurs whenever a new channel of this kind enters the window $-eV>\mu_1>\mu_2>0$.

It is straightforward to calculate $\mu_1$ and $\mu_2$ as functions of the voltages $V_G$, $V_{DS}$ and the initial numbers (m,n) of electrons, for given capacitances $C_L$, $C_M$, $C_G$.[14] From this one can determine the number of accessible current-carrying channels, and the results are shown in Fig. 3(a). There is a peak in $dI/dV_{DS}$ at the voltages at which a new channel is added, giving the set of lines in Fig. 3(b). The capacitances are adjusted to make the lines agree with the data, also shown in Fig. 3(b).

The double-diamond structure at small $V_{DS}$ clearly identifies the Coulomb blockade region of the device; the values of (n,m) for the successive diamonds is indicated. At higher $V_{DS}$, the differential conductance shows evidence of the two charge transfer processes, for electrons or holes (or both) sequentially tunneling through the device [Fig. 3(a)]. Comparing the data in Fig. 3(b) to the diagram in Fig. 3(a), one can determine the number of tunneling channels at any values of $V_{DS}$ and $V_G$.

The fitted capacitance values from Fig. 3(b), $C_L=8.2\pm0.2$ aF, $C_M=4.5\pm0.1$ aF and $C_G=0.785\pm0.015$ aF agree reasonably well with the results of our finite-element analysis, $C_L=4.9$ aF, $C_M=3.7$ aF and $C_G=0.84$ aF. The model underestimates $C_L$ and $C_M$ to some extent, probably because we omitted parallel NC facets, whereas the agreement with experiment is much better for $C_G$, for which the exact crystal shape is less significant.

A remarkable property of this device is the high ratio of gate capacitance to total capacitance, which is an order of magnitude larger than for previous NC devices.[6] This is obviously the result of incorporating a NC into the edge of each electrode, reducing the capacitance between the outermost island and the electrode. This makes it possible to access multiple charge states with moderate gate voltages, an important requirement for nanoelectronic devices. Another valuable feature is the high stability of the electronic properties of our devices, whereas previous authors reported multiple "switching" events due to changes in background charges.[6,15] This property, the origin of which is not yet fully understood, will be of practical importance for a future implementation of NC-based electronic devices.

In summary, we have demonstrated a new fabrication route for making one-dimensional metal-island arrays based on a combination of self-assembly of surfactant-coated NCs and lithography. The electronic properties of a double-island device at 4.2 K are successfully predicted with combined finite element and Monte Carlo modeling. Furthermore, all features in the differential conductance are explained by the conventional Coulomb blockade model. This fabrication method may be useful for studying other nano-electronic devices.

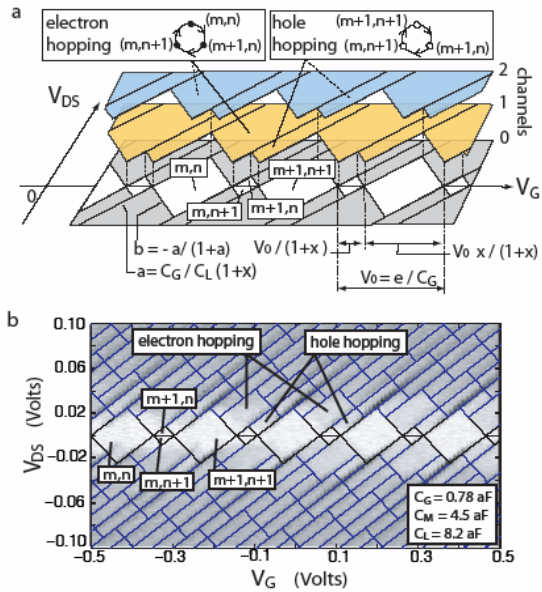

FIG. 3. Characteristics of a double-island device. (a) The number of tunneling channels accessible as a function of $V_{DS}$ and $V_G$. At $V_{DS}=0$ the numbers (m,n) of electrons on each island is constant inside the white diamonds. At higher $V_{DS}$ tunneling channels open for either electrons or holes. The conduction channels are bounded by lines, the positions and slopes (a>0 and b<0) of which only depend on the device capacitances $C_G$, $C_M$, $C_L$ and the inter-island capacitive coupling ratio $x=C_M/(C_G+C_M+C_L)$. The tunneling states for $V_{DS}<0$ are omitted for clarity. (b) Fit of the double-island differential conductance data shown in Fig.2(a) (background picture) with the conductivity jumps expected from the diagram in **a**. The fitting parameters are listed in the inset.

Acknowledgements: The authors thank G. Granger, S. Amasha, M. Gudiksen, T. Mentzel, J. Philip, A. Dorn, J. Tracy, M. Zahn and F. Stellacci for discussions and experimental support. Financial support from the German Academic Exchange Service (for D.N.W.), the NSF (Harvard NSEC) and the David and Lucile Packard Foundation is gratefully acknowledged. This work was funded in part by the NSF MRSEC program under Award Number DMR 0213282 at MIT and made use of its shared user facilities, and in part by the NSEC Program of the National Science Foundation under Award Number DMR-0117795.